\def\rfr#1{eq. (\ref{#1})}

\def\Rfr#1{Eq. (\ref{#1})}

\def\asec{$''$ cy$^{-1}$}

\def\dert#1#2{\frac{{{d}}{#1}}{{{d}}{#2}}}              

\def\asec{$''$ cy$^{-1}$}

\def\bar{\begin{eqnarray}}
\def\ear{\end{eqnarray}}
\def\bb{\bibitem}
\def\eqi{\begin{equation}}
\def\eqf{\end{equation}}
\def\eqia{\begin{eqnarray}}
\def\eqfa{\end{eqnarray}}
\def\rp#1#2{{#1\over#2}}

\def\lb#1{\label{#1}}

\def\oc2{$\mathcal{O}(c^{-2})$}

\documentclass[11pt]{article}
\usepackage{amsmath,amsthm,amscd,amssymb}
\usepackage{latexsym}
\usepackage{graphicx,epsfig}

\begin{document}

\noindent{\bf \LARGE{Dynamical determination of the Kuiper Belt's
mass from motions of the inner planets of the Solar System}}
\\
\\
\\
{Lorenzo Iorio}\\
{\it Viale Unit$\grave{a}$ di Italia 68, 70125\\Bari, Italy
\\tel./fax 0039 080 5443144
\\e-mail: lorenzo.iorio@libero.it}

\begin{abstract}
In this paper we dynamically determine the mass of the Kuiper Belt
Objects by exploiting the latest observational determinations of
the orbital motions of the inner planets of the Solar System. Our
result, in units of terrestrial masses, is $0.033\pm 0.115$ by
modelling the Classical Kuiper Belt Objects as an ecliptic ring of
finite thickness. A two-rings model yields for the Resonant Kuiper
Belt Objects a value of $0.018\pm 0.063$. Such figures are
consistent with recent determinations obtained with ground and
space-based optical techniques. Some implications for precise
tests of Einsteinian and post-Einsteinian gravity are briefly
discussed.
\end{abstract}


\section{Introduction}
Starting in 1992, astronomers have become aware of a vast
population of small bodies orbiting the Sun beyond
Neptune\footnote{See
http://www.ifa.hawaii.edu/faculty/jewitt/kb.html}. There are at
least 70,000 Trans-Neptunian Objects (TNOs) with diameters larger
than 100 km in the 30-50 AU region. Observations show that TNOs
are mostly confined within a thick band around the ecliptic,
leading to the realization that they occupy a ring or belt
surrounding the Sun. This ring is generally referred to as the
Kuiper\footnote{ See
http://www.ifa.hawaii.edu/faculty/jewitt/kb/gerard.html for the
origin of the name.}  Belt (Edgeworth 1943; Kuiper 1951; Fernandez
1980).

Reasons of interest in the Kuiper Belt are as follows
\begin{itemize}
\item
It is likely that the Kuiper Belt Objects (KBOs) are extremely
primitive remnants from the early accretional phases of the solar
system. The inner, dense parts of the pre-planetary disk condensed
into the major planets, probably within a few millions to tens of
millions of years. The outer parts were less dense, and accretion
progressed slowly. Evidently, many small objects were formed
\item
It is widely believed that the Kuiper Belt is the source of the
short-period comets. It acts as a reservoir for these bodies in
the same way that the Oort Cloud acts as a reservoir for the
long-period comets
\item
KBOs are usually not yet modelled in the data reduction softwares
used for orbit determination purposes, so that they may represent
a serious bias in precise tests of Einsteinian and
post-Einsteinian gravity. Thus, it is important to assess their
impact on planetary motions
\end{itemize}

KBOs can be classified into three dynamical classes (Jewitt et al.
1998)
\begin{itemize}
\item
Classical KBOs (CKBOs), following nearly circular orbits with
relatively low eccentricities ($e<0.25$) and semimajor axes $41\
{\rm AU}\lesssim a\lesssim 46\ {\rm AU}$; they constitute about
70$\%$ of the observed population
\item
Resonant KBOs occupy mean motion resonances with Neptune, such as
3:2 (the Plutinos, $a\sim 39.4$ AU) and 2:1 ($a\sim 47.8$ AU), and
amount to about 20$\%$ of the known objects
\item
Scattered KBOs,
which represent only about 10$\%$ of the known KBOs but have the
most extreme orbits, with $a\sim 90$ AU and $e\sim 0.6$,
presumably due to a weak interaction with Neptune; we only have
rather poor knowledge of them.
\end{itemize}
In this paper we determine the mass of  KBOs in a truly dynamical
way by means of the latest observational determinations of the
secular perihelion advances of the inner planets of the Solar
System. The use of such celestial bodies presents many advantages.
Long data sets including also many accurate radio-technical range
and range-rate measurements are available; they cover many orbital
revolutions, contrary to the outer planets which are more affected
by KBOs like Uranus and Neptune. Indeed, mainly optical data and
sparse radar-ranging measurements exist for them covering barely
one period for Uranus and less than one full orbital revolution
for Neptune. Moreover, for heliocentric distances of the order of
just 1 AU or less many details of the true mass distribution of
KBOs are not relevant and all the recently proposed analytical
models converge satisfactorily to a substantially unified
description, given the present-day accuracy in reconstructing the
orbital motions in our region of the Solar System. This fact
greatly simplifies the analytical calculation and allows for
reliable, rather model-independent determinations of the KBOs's
mass.

\section{The determined extra-rates of perihelion}
 The Russian astronomer E.V. Pitjeva
(Institute of Applied Astronomy, Russian Academy of Sciences)
recently processed almost one century of data of all types in the
effort of continuously improving the EPM2004 planetary ephemerides
(Pitjeva 2005a). Among other things, she also determined residual
secular, i.e. averaged over one orbital revolution, rates of the
longitudes of perihelion $\varpi=\Omega+\omega$, where $\omega$
and $\Omega$ are the argument of perihelion and the longitude of
the ascending node, respectively, of the inner  planets (Pitjeva
2005b) as fit-for parameters of global solutions in which she
contrasted, in a least-square way, the observations (ranges,
range-rates, angles like right ascension $\alpha$ and declination
$\delta$, etc.) to their predicted values computed with a complete
suite of dynamical force models. The results are shown in Table
\ref{pitabl}.
{\small\begin{table}\caption{ Observationally determined
extra-precessions of the longitudes of perihelia of the inner
planets, in arcseconds per century (\asec), by using EPM2004 with
$\beta=\gamma=1$, $J_2=2\times 10^{-7}$, from Table 3 of (Pitjeva
2005b). Both KBOs and the general relativistic gravitomagnetic
force were not included in the adopted dynamical force models. The
quoted uncertainties are not the mere formal, statistical errors
but are realistic in the sense that they were obtained from
comparison of many different solutions with different sets of
parameters and observations. The correlations among such
determined planetary perihelia rates are very low with a maximum
of about $20\%$ between Mercury and the Earth (Pitjeva, private
communication 2005). }\label{pitabl}

\begin{tabular}{llll} \noalign{\hrule height 1.5pt}

 Mercury & Venus  & Earth & Mars\\
$-0.0036\pm 0.0050$ & $0.53\pm 0.30$ & $-0.0002\pm 0.0004$ & $0.0001\pm 0.0005$\\
\hline

\noalign{\hrule height 1.5pt}

\end{tabular}

\end{table}}
The modelled forces are
\begin{itemize}
\item
The Newtonian N-body perturbations including also the effect of
301 largest asteroids (Krasinsky et al. 2002) and of the minor
asteroid ring in the ecliptic
\item
The Sun's quadrupole mass moment $J_2$ (Patern\`{o} et al. 1996;
Pijpers 1998; Mecheri et al. 2004), set to $2\times 10^{-7}$
\item
The post-Newtonian gravitoelectric forces (Newhall et al. 1983)
parameterized in terms of the PPN Eddington-Robertson-Schiff
parameters $\gamma$ and $\beta$ (Will 1993) set to their general
relativistic values $\gamma=\beta=1$
\end{itemize}
The un-modelled forces are
\begin{itemize}
\item
The Newtonian force induced by KBOs
\item The post-Newtonian gravitomagnetic force responsible
for the Lense-Thirring effect (Lense and Thirring 1918).
\end{itemize}
Thus, the effect of KBOs is entirely accounted for by the
so-obtained residual perihelia advances of Table \ref{pitabl}.

\section{Modelling and confrontation with data}
In order to make a comparison with Table \ref{pitabl}, a
theoretical prediction for the secular precessions induced by KBOs
on the planetary perihelia is required.

The  action of KBOs can be treated in a perturbative way. The
Gauss rate equations for $\omega$ and $\Omega$ are, for a generic
orbital configuration \eqi\dert\omega t=-\cos i\dert\Omega
t+\rp{\sqrt{1-e^2}}{nae}\left[-A_r\cos
f+A_t\left(1+\rp{r}{p}\right)\sin f\right],\lb{pri}\eqf and
\eqi\dert\Omega t=\rp{A_n}{na\sqrt{1-e^2}\sin i
}\left(\rp{r}{a}\right)\sin(\omega+f),\lb{nd}\eqf where $i$ is the
the inclination of the orbit, $p=a(1-e^2)$, $n=\sqrt{GM/a^3}$ is
the Keplerian mean motion, $f$ is the true anomaly counted from
the pericentre, and $A_r$, $A_t$ and $A_n$ are the radial,
transverse and normal components of the perturbing acceleration
$\boldsymbol A$, respectively.

For almost ecliptic orbits ($i\sim 0$ deg, $\cos i\sim 1$, $\sin
i\sim i$), like those of the Solar System's major bodies, the rate
equation for $\varpi$ can safely be approximated as
\eqi\dert\varpi t\sim \rp{\sqrt{1-e^2}}{nae}\left[-A_r\cos
f+A_t\left(1+\rp{r}{p}\right)\sin f\right].\lb{mary}\eqf Its
secular rate
is obtained by evaluating the right-hand-side of \rfr{mary} onto
the unperturbed Keplerian ellipse, characterized in terms of the
eccentric anomaly $E$ by
\begin{equation}\left\{\begin{array}{lll}
r=a(1-e\cos E),\\\\
\cos f=\rp{\cos E-e}{1-e\cos E
},\\\\
\sin f=\rp{\sin E\sqrt{1-e^2}}{1-e\cos E},
\lb{eccen}\end{array}\right.\end{equation}
and averaging the result over one orbital period $P$ by means of
\eqi \rp{dt}{P}=\rp{(1-e\cos E)}{2\pi}dE. \lb{prd}\eqf



Motivated by the search for a gravitational explanation of the
Pioneer anomaly (Anderson et al. 2002), various authors recently
produced several analytical models for the KBOs acceleration
(Anderson et al. 2002; Nieto 2005; Bertolami and Vieira 2006; de
Diego et al. 2006). At heliocentric distances $\lesssim 1$ AU many
details of the three-dimensional mass distribution of KBOs can be
neglected, so that we can approximate it with a bi-dimensional
distribution lying in the ecliptic plane. Moreover, the four
models analyzed by Bertolami and Vieira (2006), i.e. two-rings,
uniform disk, non-uniform disk and torus, give the same results
for $r\ll 20$ AU and for various ecliptic latitudes $\beta$ close
to zero. Thus, we will adopt for CKBOs the uniform disk model,
which consists of a uniform, hollow thin disk lying in the
ecliptic plane within distances $R_{\rm min}=37.8$ AU and $R_{\rm
max}=46.2$ AU (Bernstein et al. 2004).

The Newtonian gravitational field of a thin massive ring exhibits
cylindrical symmetry and has an in-plane, radial component and an
out-of-plane, normal component (Owen 2003). Since we are looking
for the effects induced on the longitude of the perihelion, only
the radial component is relevant to us. Thus, from
\rfr{mary}$-$\rfr{prd} the secular perihelion rate can be written
as \eqi\left\langle\dert\varpi t\right\rangle=-\rp{1}{2\pi
e}\sqrt{\rp{a(1-e^2)}{GM}}\int_0^{2\pi}A_r(E)(\cos
E-e)dE.\lb{perirate}\eqf The radial component of $\boldsymbol
A_{\rm K}$ for a uniform disk is (Bertolami and Vieira 2006) \eqi
A_{\rm ud}=\mathcal{K}\int_0^{2\pi}\int_{R_{\rm min}}^{R_{\rm
max}} \rp{R_{\rm K}\left[r-R_{\rm K}\cos\beta\cos(\phi_{\rm
K}-\lambda)\right]}{\left[R_{\rm K}^2+r^2-2rR_{\rm K
}\cos\beta\cos(\phi_{\rm K}-\lambda)\right]^{3/2}}d\phi_{\rm
K}dR_{\rm K}, \lb{accel2r}\eqf
%
where \eqi\mathcal{K}=-\rp{Gm_{\rm K}}{2\pi(R_{\rm max}^2-R_{\rm
min}^2)},\eqf $m_{\rm K}$ is the KBOs's mass, $R_{\rm K}$ and
$\phi_{\rm K}$ are the polar coordinates of the disk mass element,
and $\{r,\beta,\lambda\}$ are the usual spherical ecliptic
planetary coordinates. For the inner planets, with $r\ll R_{\rm
min}, R_{\rm max}$ and whose orbits lie almost exactly in the
ecliptic plane, we can safely pose\footnote{Recall that
$\cos\beta=\sqrt{1-\sin^2 i\sin^2 f}$. }
\eqi\left[1+\left(\rp{r}{R_{\rm K }}\right)^2-2\left(\rp{r}{R_{\rm
K }}\right)\cos\beta\cos(\phi_{\rm K }-\lambda)\right]^{-3/2}\sim
1+3\left(\rp{r}{R_{\rm K}}\right)\cos(\phi_{\rm
K}-\lambda).\lb{approx}\eqf By inserting \rfr{approx} into
\rfr{accel2r} and performing the integration we get
\eqi A_{\rm ud}=\rp{Gm_{\rm K}}{2(R_{\rm max}+R_{\rm min})R_{\rm
max}R_{\rm min }}r.\lb{accfin}\eqf Note that it is positive, in
according with the behavior of all the models considered in
(Bertolami and Vieira 2006) for $r\ll 20$ AU.
\Rfr{accfin}, with $r=a(1-e\cos E)$, into \rfr{perirate} finally
yields \eqi\left\langle\dert\varpi
t\right\rangle=\rp{3}{4}\sqrt{\rp{Ga^3(1-e^2)}{M}}\rp{m_{\rm
K}}{(R_{\rm max}+R_{\rm min })R_{\rm max}R_{\rm
min}}.\lb{finale}\eqf
%
%

Table \ref{pitabl} and \rfr{finale} can now be used to determine
$m_{\rm K}$ in a truly dynamically, model-independent way. To this
aim, let us note that, by construction, the determined extra-rates
of Table \ref{pitabl} are not only due to KBOs, but also the
Lense-Thirring field and the mismodelled part of the solar $J_2$,
which is presently uncertain at a $\sim 10\%$ level, contribute to
them. Their nominal magnitudes are listed in Table \ref{solgm} and
Table \ref{solj2}, respectively.
\begin{table}\caption{Gravitomagnetic secular precessions of the
longitudes of  perihelion $\varpi$ of Mercury, Venus, Earth and
Mars in \asec. The value $(190.0\pm 1.5)\times 10^{39}$ kg m$^2$
s$^{-1}$ (Pijpers 1998; 2003) has been adopted for the solar
proper angular momentum $L_{\odot}$. }\label{solgm}

\begin{tabular}{llll}
\noalign{\hrule height 1.5pt}

 Mercury & Venus  & Earth & Mars\\
$-0.0020$ & $-0.0003$ & $-0.0001$ & $-0.00003$\\ \hline

\noalign{\hrule height 1.5pt}
\end{tabular}
\end{table}
\begin{table}\caption{ Nominal values of the classical secular
precessions of the longitudes of  perihelion $\varpi$ of Mercury,
Venus, Earth and Mars, in \asec, induced by the solar quadrupolar
mass moment $J_2$. The value $J_2=2\times 10^{-7}$ used in
(Pitjeva 2005b) has been adopted. Their mismodelled amplitudes can
be obtained  by assuming an uncertainty in $J_2$ of the order of
$\sim 10\%$.}\label{solj2}

\begin{tabular}{llll}
\noalign{\hrule height 1.5pt}

 Mercury & Venus  & Earth & Mars\\
0.0254 & 0.0026 & 0.0008 & 0.0002 \\ \hline

\noalign{\hrule height 1.5pt}
\end{tabular}

\end{table}
Although the Lense-Thirring effect and the mismodelled part of the
$J_2$ precessions may be neglected in our analysis, being smaller
than the errors of Table \ref{pitabl}, we prefer conservatively to
cancel out, by construction, any possible impact due to them. It
can be done by suitably combining the perihelia of Mercury, the
Earth and Mars according to an approach followed, e.g., in (Iorio
2005). By converting Table \ref{pitabl} in s$^{-1}$, we have the
CKBOs' mass, in units of terrestrial masses \eqi m^{(\rm
comb)}_{\rm K }=\rp{\dot\varpi_{\rm Mercury}+c_1\dot\varpi_{\rm
Earth}+c_2\dot\varpi_{\rm Mars } }{1.017\times 10^{-15}\ {\rm
s}^{-1}}=0.052,\lb{massak}\eqf with \eqi c_1=-81.71,\
c_2=221.18.\eqf Such coefficients, built in terms of the semimajor
axes and eccentricities of the involved planets, assure that the
solar Newtonian quadrupolar and post-Newtonian gravitomagnetic
fields, whatever their contributions to  Table \ref{pitabl} is, do
not affect at all the recovered CKBOs' mass. Because of the
existing correlations among the determined extra-rates of
perihelia, the error can be conservatively evaluated as \eqi
\delta m^{(\rm comb)}_{\rm K }\leq\rp{\delta\dot\varpi_{\rm
Mercury}+|c_1|\delta\dot\varpi_{\rm
Earth}+|c_2|\delta\dot\varpi_{\rm Mars } }{1.017\times 10^{-15}\
{\rm s}^{-1}}=0.223.\eqf

If we only use the Mars perihelion, whose Lense-Thirring and
mismodelled $J_2$ precessions are negligible by one order of
magnitude, we get \eqi m_{\rm K}^{(\rm Mars )}=0.026\pm
0.134.\lb{marsonly}\eqf

A weighted mean of such two measurements yields \eqi m_{\rm
K}^{(\rm weighted )}=0.033\pm 0.115.\lb{wei}\eqf

In regard to the Resonant KBOs, we can adopt the two-rings model
with $R_1=39.4$ AU and $R_2=49.8$ AU. The radial acceleration is
(Bertolami and Vieira 2006) \eqi A_{\rm 2R}=-\rp{Gm_{\rm
K}}{2\pi(R_1+R_2)}\int_0^{2\pi}\sum_{i=1}^2
\rp{R_i\left[r-R_i\cos\beta\cos(\phi_{\rm
K}-\lambda)\right]}{\left[r^2+R_i^2-2rR_i\cos\beta\cos(\phi_{\rm
K}-\lambda)\right]^{3/2}}d\phi_{\rm K}. \lb{2ring}\eqf By using
the same approximation of \rfr{approx}, we obtain $m_{\rm K}^{(\rm
comb)}=0.029\pm 0.123$ and $m_{\rm K}^{(\rm Mars)}=0.015\pm
0.073$, with $m_{\rm K}^{(\rm weighted )}=0.018\pm 0.063$. Note
that such figures are smaller than those for CKBOs, consistently
with the existing estimates for the population of Resonant KBOs.

At this point, we can {\it a posteriori} justify the use of the
approximation of \rfr{approx} for both the models used: indeed, by
using our values for $m_{\rm K}$ it is possible to show that
additional terms in the expansion of \rfr{approx} yield
precessions far too small to be detected with the present-day
accuracy.

\section{Discussion and conclusions}
In this paper we dynamically determined the mass of KBOs from an
analysis of the recently determined secular perihelion advances of
the rocky planets of the Solar System. We modelled CKBOs as a
uniform thick ring; such a simple model is justified by the fact
that at heliocentric distances of about 1 AU many details of the
true three-dimensional KBOs mass distribution can be neglected;
indeed, many different, more or less complicated analytical models
manifest substantially the same behavior at distances much less
than 20 AU. For CKBOs we obtained a mass of $0.033\pm 0.115$, in
units of terrestrial masses. A two-rings model for the Resonant
KBOs yields a mass of $0.018\pm 0.063$.

Such figures are consistent with those by
\begin{itemize}
\item
Bernstein et al. (2004), who give a nominal mass of $0.010$ for
CKBOs and $0.021$ for their Excited class including some Plutinos
and Scattered KBOs. They used the ACS camera of the Hubble Space
Telescope
\item
Gladman et al. (2001) yielding a mass of $0.04-0.1$ for all the
TNOs in the range 30-50 AU, excluding the Scattered KBOs. The
adopted technique was deep imaging on the Canada-France-Hawaii
Telescope and the ESO Very Large Telescope UT1
\item
 Trujillo et al. (2001) giving for CKBOs
a value of 0.030, from a wide-field survey with the CCD Mosaic
camera of the Canada-France-Hawaii Telescope.
\end{itemize}
The upper limit of 0.3 terrestrial masses obtained by Backman et
al. (1995) from far-IR emission measurements is, instead, ruled
out.

It is important to note that all the previous estimates  are based
on various assumptions about, e.g., albedo and density of the KBOs
for which very large uncertainties still exist, so that the
authors of the previously cited works decided to release no errors
of their mass estimations.

The impact of KBOs on the inner planets of the Solar System lies
at the edge of the present-day accuracy. Its induced effects,
shown in Table \ref{kbtab} for a particular value of the CKBOs'
mass,
\begin{table}\caption{ Nominal values of the classical secular
precessions of the longitudes of  perihelion $\varpi$ of Mercury,
Venus, Earth and Mars, in \asec, induced by CKBOs. For $m_{\rm K
}$ the value of \rfr{massak} has been used. }\label{kbtab}

\begin{tabular}{llll}
\noalign{\hrule height 1.5pt}

 Mercury & Venus  & Earth & Mars\\
$0.00002\pm 0.0001$ & $0.00006\pm 0.0003$ & $0.0001\pm 0.0004$ & $0.0002\pm 0.0008$   \\
\hline

\noalign{\hrule height 1.5pt}
\end{tabular}

\end{table}
should not be neglected in, e.g., precision tests of gravity
because, especially for the Earth and Mars,  they are of the same
order of magnitude, or even larger, than some Einsteinian (see
Table \ref{solgm}) and post-Einsteinian features of
motion\footnote{In the context of the multidimensional model by
Dvali, Gabadadze and Porrati (Dvali et al. 2000), secular
perihelion precessions of the order of 0.0005 \asec\ are predicted
for the Solar System planets (Lue and Starkman 2003; Iorio
2005b).} which recently attracted much attention in view of a
possible detection in the near future. If not accounted for in the
dynamical force models of the orbit data reduction softwares, KBOs
may bias the recovery of such effects when the required precision
level will be finally attained. In the case of the measurement of
the Lense-Thirring effect by only using the perihelion of Mercury
(Iorio 2005a), the impact of KBOs is negligible. It is not so if a
combination of the perihelia of Mercury and the Earth is used in
order to cancel out the mismodelling of $J_2$ (Iorio 2005a).
Thanks to the availability of a simple and reliable formula as
that of \rfr{finale} for the inner planets, such a problem could
be circumvented by setting a suitable three-elements combination
allowing for de-coupling the Lense-Thirring effect from $J_2$ and
KBOs as well.

Incidentally, let us note that our results further enforce the
conclusion that the Pioneer anomaly cannot be due to KBOs
(Anderson et al. 2002; Nieto 2005; Bertolami and Vieira 2006).


\end{document}